\documentclass[12pt]{article}
\usepackage[dvips]{graphicx,color}
\usepackage{latexsym}

\newcommand{\be}{\begin{equation}}
\newcommand{\ee}{\end{equation}}
\newcommand{\bea}{\begin{eqnarray}}
\newcommand{\eea}{\end{eqnarray}}
\newcommand{\ds}{\displaystyle}

\begin{document}

\title{Instabilities of infinite matter with effective Skyrme-type
  interactions}

\author{J. Margueron \\
GANIL, CEA/DSM - CNRS/IN2P3 \\ B.P. 5027, F-14076 Caen CEDEX 05, France\\
\\
J. Navarro \\
IFIC (CSIC - Universidad de Valencia) \\ Apdo. 22085, 
E-46.071-Valencia, Spain \\
\\
Nguyen Van Giai \\
Institut de Physique Nucl\'eaire, Universit\'e Paris Sud \\ F-91406 Orsay CEDEX,
France}

\maketitle

\abstract{The stability of the equation of state predicted by Skyrme-type
  interactions is examined. We consider simultaneously symmetric nuclear
  matter and pure neutron matter. The stability is defined by the
  inequalities that the Landau parameters must satisfy simultaneously. A
  systematic study is carried out to define interaction parameter domains
  where the inequalities are fulfilled. It is found that there is always a
  critical density $\rho_{cr}$ beyond which the system becomes unstable. The
  results indicate in which parameter regions one can find effective forces
  to describe correctly finite nuclei and give at the same time a
  stable equation of state   up to densities of $3-4$ times the saturation 
  density of symmetric nuclear matter.}

\vspace{1cm}
\section{Introduction}

Nucleon-nucleon effective interactions with an explicit density dependence 
have been largely employed since the seventies for studies of nuclear
properties. Once the form of the interaction is chosen, either a 
zero-range force of Skyrme-type~\cite{vau72} or a finite-range force of 
Gogny-type~\cite{gog75}, the parameters are determined by fitting some 
selected properties of doubly-magic nuclei and symmetric nuclear matter at 
saturation. Within a Hartree-Fock (HF) scheme these interactions are able to 
describe quantitatively the properties of stable as well as unstable exotic 
nuclei. The interactions are thus well controlled around the saturation 
density $\rho_0$ of symmetric nuclear matter, for moderate isospin asymmetries 
and zero temperature. This type of effective interactions has also been 
employed to study nuclear matter in conditions of astrophysical interest, 
for instance neutron matter with a finite proton fraction at densities up to 
several times $\rho_0$ and finite temperatures. This system at such conditions 
is relevant for studying proto-neutron stars. The effective interactions 
are thus extrapolated to conditions of density and isospin asymmetry 
which are not experimentally accessible, and one should then ask for 
the limits of validity of such an extrapolation. 

Of course, the description of nuclear matter consisting of nucleons 
is no longer valid beyond some value of the density, as other degrees
of freedom appear. For instance, strange matter is expected beyond 
$3-4\rho_0$~\cite{sch00}. For the present discussion, we shall arbitrarily
accept that for densities up to $4\rho_0$, only nucleons are needed to 
describe nuclear matter. Even in this range of densities the effective 
interactions are not well determined. For instance, different effective 
interactions can give similar equations of state for symmetric nuclear 
matter and very different results for pure neutron matter. Indeed, the 
extreme asymmetry of isospin is not part of the usual input to determine the 
interaction parameters. In some cases, the determination of the parameters 
includes a fit to the equation of state of neutron matter calculated 
with microscopic methods and realistic interactions. The Skyrme 
interaction RATP~\cite{ray82} takes into account the variational
microscopic results of Ref.~\cite{fri81}. More recently, the set of 
SLy interactions~\cite{cha97,cha98} have included the variational
results of Ref.~\cite{wir88} among the conditions required to fix the 
parameters. It is worth noting that these microscopic variational calculations 
are not exempt from uncertainties since realistic interactions are not 
completely known, especially their three-body part. Indeed, the same 
theoretical model~\cite{wir88} employing different realistic interactions 
leads to different equations of state, both in symmetric nuclear matter 
and in pure neutron matter, and sizeable differences appear for densities 
beyond the saturation value $\rho_0$. Although this type of approach gives a 
precious guide for the determination of the effective interaction parameters, 
it cannot replace the empirical data which are the natural input for the 
phenomenological interactions. 

The extrapolation of effective interactions can result in an unphysical 
behavior of nuclear matter~\cite{mar01}. For instance, most Skyrme 
parametrizations predict that the isospin asymmetry energy $\epsilon_I$ 
becomes negative when the density is increased. Consequently, the symmetric 
system would be unstable beyond the saturation density, preferring a largely 
asymmetric system made by an excess of either protons or neutrons. Another type 
of instability refers to the magnetic properties of neutron matter. 
The possibility of a ferromagnetic transition at high densities has 
been studied long ago, employing different theoretical ingredients, and the 
results were contradictory. Currently used Skyrme interactions predict that 
neutron matter becomes spin polarized at densities between 
$\simeq 1.1-3.5\rho_0$~\cite{mar01,vid84}. This transition would have 
important consequences for the evolution of a proto-neutron star since the 
mean free path of neutrinos would be zero~\cite{mar01,nav99}. However, 
Gogny-type effective interactions either exclude such a ferromagnetic 
transition or predict it at very high densities~\cite{mar01}. Relativistic 
mean-field calculations~\cite{nie90,mar91,ber95} predict that such a transition 
could appear at densities beyond $\simeq 4\rho_0$. Finally, 
recent Monte Carlo simulations~\cite{fan01} as well as Brueckner-Hartree-Fock 
calculations~\cite{polls,lombardo} using modern two- and three-body realistic 
interactions do exclude such an instability. 

The purpose of this work is to analyze whether the presence of
instabilities is inherent or not to the Skyrme-type interactions.
In Section 2 we explain the restrictions on the Skyrme parameters
imposed by the requirement that no instabilities should appear at
densities in the range $(1-4)\rho_0$. Furthermore, the sound velocity
$v_s/c$ in matter must remain smaller than unity, and this adds a new
constraint. Section 3 contains the discussion of
the results. Conclusions are drawn in Section 4. Useful
expressions are given in Appendix A and B.

\section{Constraints from the Landau parameters}

The condition that the spin-unpolarized neutron matter should be the most
stable phase at any density was used in Ref.~\cite{kut94} to constrain 
the Skyrme parameters. As the potential energy of the fully spin-polarized
phase only depends on the combination $t_2 ( 1 + x_2)$, ferromagnetic collapse 
is necessarily avoided if this combination is positive. Taking into account
current Skyrme parametrizations, the authors of Ref.~\cite{kut94} concluded 
that
$-5/4<x_2<-1$. However, this condition is not sufficient: the family of SLy
parametrizations~\cite{cha97,cha98} imposes this constraint but 
at densities of about 0.37~fm$^{-3}$, a ferromagnetism instability is
predicted in neutron matter~\cite{mar01}. 

Here, we make a systematic use of the stability criteria related to the 
adimensional Landau parameters $F_l$, $G_l$, $F'_l$ and $G'_l$ in symmetric 
nuclear matter, and $F^{(n)}_l$ and $G^{(n)}_l$ in neutron matter. 
These criteria simply establish that any parameter of multipolarity 
$l$ should be greater than $-(2l+1)$. Current effective interactions 
satisfy these conditions at saturation, and we shall investigate 
if they can be maintained for densities up to $4\rho_0$.

The general form of the Skyrme interaction contains ten parameters,
eight of them denoted as $t_i$, $x_i$ (with $i=0,1,2,3$), the other
two being the power $\sigma$ of the density dependence and the
strength $W_0$ of the zero-range spin-orbit term. These parameters
are usually determined by fitting some selected properties of finite nuclei 
and symmetric nuclear matter. In this study we follow a different 
procedure, using some accepted nuclear matter values to fix as many 
interaction parameters as possible and using the stability criteria
to put bounds on the remaining parameters.

Our starting point is given by the empirical
values of the following quantities
in symmetric nuclear matter at the saturation point: density $\rho_0$,
energy per particle $\epsilon_0$, effective mass $m^*_0$,
compression modulus $K_0$, surface energy $\epsilon_S$, and
isospin asymmetry energy $\epsilon_I$. Imposing that a Skyrme-type
interaction reproduces the values of $\rho_0$, $\epsilon_0$, $m^*_0$,
and $K_0$ completely determines the parameters $t_0$, $t_3$, $\sigma$
and the combination $T_0 \equiv [3t_1+(5+4x_2)t_2]/8$. The explicit
expressions are given in Appendix A. With a Skyrme interaction, the
surface energy $\epsilon_S$ depends on the previously fixed parameters, on
the combination $T_S \equiv [9 t_1 - (5+4x_2)t_2]/8$ and on the
strength $W_0$ of the spin-orbit term~\cite{trekri}. The value of
$W_0$ is usually determined by fitting the spin-orbit splitting of some 
selected levels in finite nuclei, and for the present discussion we shall
assume the often used value $W_0 = 120$~MeV$\cdot$fm$^5$. Furthermore, the 
symmetry energy $\epsilon_I$ can be expressed in terms of the above parameters
and of the four combinations $t_i x_i$ ($i=0 - 3$). In summary, from
the ten Skyrme parameters, one is kept fixed ($W_0$), six parameters or
combinations are determined by the six empirical inputs, and three
combinations are free. We found convenient to choose the free combinations
$x \equiv t_1x_1$, $y \equiv t_2x_2$, $z \equiv t_3x_3$. 

From here on, our task is to analyze the possible domains of the 
$(x,y,z)$-space for different densities. For a given value of the density 
$\rho$ we would like to know what is the $(x,y,z)$ domain where no instability 
can occur. Stability implies that any 
adimensional Landau parameter of multipolarity $l$ 
must be larger that $-(2l+1)$. Skyrme forces only contain monopolar
and dipolar contributions to the particle-hole interaction so that
all Landau parameters are zero for $l > 1$. Thus, we have twelve
inequalities,  eight coming from symmetric nuclear matter, and four from 
neutron matter. Explicit expressions of the twelve Landau parameters are 
given in Appendix B.  

Two of the twelve inequalities, however, play a special role.
The condition $F_1 > -3$ is trivially satisfied for any density
as long as $ 0 < m^*(\rho_0) <m$, which is the case.  
The Landau parameter $F_0$ only depends on $t_1$, 
$t_3$, $\sigma$ and the combination $T_0$, which are determined by 
the initial inputs. Exploring the inequality $F_0 > -1$ as a function 
of the density, it is found that it is violated at a density smaller than  
$ \rho_0$. It corresponds to a density where the compression 
modulus becomes negative, which is the spinodal point, i.e, the
occurrence of a  liquid-gas transition. Note that within a Skyrme 
interaction framework, the spinodal density is completely determined
by the saturation properties $\epsilon_0$, $\rho_0$, $K_0$ and $m^*_0$.
After excluding the inequalities related to $F_1$ and $F_0$, we are left with 
ten inequalities to be satisfied by the yet free combinations $x,y,z$.

\subsection{Symmetric nuclear matter}

The parameters $G'_0$ and $G'_1$ give two constraints for $y$:
\bea
y &<& - \frac{10 C_1(\rho)}{3 \alpha_1 \rho^{2/3}}
+ \frac{2}{3} ( T_0 - 2 T_S)~, \label{gp0} \\
y &>& - \frac{10 C_0(\rho)}{\rho} +
\frac{2}{3} ( T_0 - 2 T_S )~, \label{gp1}
\eea
where the $C_i$ and $\alpha_i$ are defined in Appendix A. 
The parameters $F'_1$, $G_1$ give two constraints on combinations of
$x$ and $y$:
\bea
x - \frac{3}{5} y &>& - \frac{4 C_0(\rho)}{\rho}
+ \frac{4}{15} ( T_0 - 2 T_S )~, \label{fp1} \\
x + \frac{3}{5} y &<& \frac{4 C_0(\rho)}{\rho}
- \frac{4}{15} ( T_0 - 2 T_S )~. \label{g1} 
\eea
The remaining Landau parameters $G_0$ and $F'_0$ constrain combinations
of all three $x$, $y$ and $z$ unknowns:
\bea 
&& \frac{1}{9 \alpha_1} ( \rho^{\sigma} - \rho_0^{\sigma})
z + ( \rho^{2/3} - \rho_0^{2/3}) x + \frac{3}{5}
( \rho^{2/3} + \rho_0^{2/3}) y > \label{g0} \\
&&\hspace{2cm} \frac{4}{3 \alpha_1} \left( C_1(\rho) + C_1(\rho_0) +
\frac{2 \epsilon_I}{\rho_0} \right) -
\frac{4}{15} ( \rho^{2/3} + \rho_0^{2/3}) ( T_0 - 2 T_S)~,
\nonumber \\
&& - \frac{1}{9 \alpha_1} ( \rho^{\sigma} - \rho_0^{\sigma})
z - ( \rho^{2/3} - \rho_0^{2/3}) x + \frac{3}{5}
( \rho^{2/3} - \rho_0^{2/3}) y > \label{fp0} \\
&&\hspace{2cm} \frac{4}{3 \alpha_1} \left( C_1(\rho) - C_1(\rho_0) -
\frac{2 \epsilon_I}{\rho_0} \right) -
\frac{4}{15} ( \rho^{2/3} - \rho_0^{2/3}) ( T_0 - 2 T_S)~.
\nonumber 
\eea
However, it is possible to take the linear combination $F'_0+G_0$ to get 
another constraint independent of $z$. Note that Eq.~(\ref{fp0}) reduces to
the trivial condition $\epsilon_I >0$ when $\rho=\rho_0$. It is due to the
fact that the symmetry energy $\epsilon_I$ has been used to write 
$x_0$ in terms of the remaining parameters.

\subsection{Neutron matter}

The parameters $F^{(n)}_1$, and $G^{(n)}_1$ give two
constraints on combinations of $x$ and $y$:
\bea
x - \frac{3}{5} y &<& \frac{4 C_0(\rho)}{\rho}
+ \frac{4}{15} ( T_0 - 2 T_S )~, \label{fn1} \\
x - \frac{1}{5} y &<& \frac{2 C_0(\rho)}{\rho}
- \frac{2}{15} ( T_0 - 2 T_S )~. \label{gn1}
\eea
The remaining Landau parameters $F^{(n)}_0$, and $G^{(n)}_0$ 
constrain combinations of all three $x,y,z$ parameters:
\bea
&& \frac{(\sigma+1)(\sigma+2) \rho^{\sigma} -
2 \rho_0^{\sigma}}{6(4 \alpha_2 \rho^{2/3} - 3 \alpha_1 \rho_0^{2/3})}
z + x  - \frac{3}{5} y < \label{fn0} \\
&&\hspace{2cm} \frac{4}{4 \alpha_2 \rho^{2/3} -3 \alpha_1 \rho_0^{2/3}}
 \left( C_1(\rho_0) - C_2(\rho) + \frac{2 \epsilon_I}{\rho_0} \right)
+ \frac{4}{15} ( T_0 - 2 T_S)~, \nonumber \\
&& \frac{1}{9} ( \rho^{\sigma} - \rho_0^{\sigma}) z +
\frac{1}{3} (2 \alpha_2 \rho^{2/3} - 3 \alpha_1 \rho_0^{2/3} ) x
+
\frac{1}{5} ( 2 \alpha_2 \rho^{2/3} + 3 \alpha_1 \rho_0^{2/3} ) y
> \label{gn0} \\
&&\hspace{2cm} \frac{4}{3} \left( C_1(\rho_0) + C_3(\rho) +
\frac{2 \epsilon_I}{\rho_0} \right) - 
\frac{4}{45} ( 7 \alpha_2 \rho^{2/3} + 3 \alpha_1 \rho_0^{2/3} )
 ( T_0 - 2 T_S)~. \nonumber
\eea
Note that the linear combination $\alpha_1 F'_0 + G^{(n)}_0$ gives 
another constraint independent of $z$. 

\subsection{Sound velocity constraint}

In addition to the constraints from stability requirements it is important
to check that for each density the sound velocity $v_s$ remains smaller than
the speed of light, i.e., superluminosity does not occur. The sound velocity
is directly related to the 
compression modulus $K(\rho)$ which can itself be expressed in terms of the
Landau parameters $F_0$ and $F_1$:
\bea
m v_s^2 & = & \frac{1}{9} K 
\label{sound1} \\
 & = & \frac{\hbar^2 k_F^2}{3 m} \frac{1 + F_0}{1 + \frac{1}{3}F_1}~.
 \nonumber  
\eea

In symmetric nuclear matter it turns out that $v_s$ depends only on the
parameters $t_0$, $t_3$, $\sigma$ and the combination $T_0$, all of which
being already determined by the six initial empirical inputs. Therefore,
there will be no additional bounds on the allowed volume in $(x,y,z)$-space.
However, Eq.~(\ref{sound1}) shows that $v_s$ increases with increasing
density and for a given choice of initial empirical inputs there is always a
value $\rho_{sound}$ beyond which $v_s/c$ is greater than 1. The value of
$\rho_{sound}$ depends essentially on the adopted value for $K_0$. For
instance, $\rho_{sound}$ is around $3.5\rho_{0}$ if $K_0$ = 350 MeV and it
may become $6\rho_{0}$ if $K_0$ = 250 MeV. Thus, there is no Skyrme
interaction which can be reasonably be used beyond $\rho_{sound}$ because it
would predict unphysical values of the sound velocity.

The situation is somewhat different in pure neutron matter. Now, the Landau
parameters in Eq.~(\ref{sound1}) depend also on the $(x,y,z)$ parameter
combinations. 
The requirement that $v_s/c$ remains smaller than unity adds one more
constraint to the determination of the allowed volume in the $(x,y,z)$
space. We shall see in the next section that the sound velocity in neutron
matter does not bring any effective restriction on the allowed volume in 
the parameter space, at least around the values of the six empirical inputs 
used in the present study. 

We must mention that a study of the sound velocity in asymmetric nuclear
matter was made in Ref.~\cite{kuo88}. The aim was to find the conditions to
be satisfied by the Skyrme parameters so that superluminosity would never
appear at any density whatsoever. The conditions which were found are very
restrictive. In the present work, our point of view is quite different since
we don't think that the Skyrme effective approach should be valid at
densities beyond $4\rho_0$. 

\section{Results and discussion}

The inputs used in the present study are:
$\epsilon_0 = -16.0$~MeV, $\rho_0 = 0.16$~fm$^{-3}$, 
$K_0 = 230$~MeV, $m^*_0/m = 0.70$, $\epsilon_S = 18.0$~MeV, 
and $\epsilon_I = 32.0$~MeV. 

For a given value of the density $\rho$, the various constraints define
in the ($x, y, z$) parameter space an allowed volume that we call
$\Omega$. The surface of $\Omega(\rho)$ is made of planes because all 
constraints are linear in $x, y, z$. Any point outside $\Omega$ is
forbidden because some of the inequalities would not be fulfilled. 
To present the results and facilitate the discussion we will
explore the $z$-axis. The intersection of a volume $\Omega(\rho)$ with a
$z$~=~constant horizontal plane gives a polygon which can be represented in the
($x,y$)-plane. By varying $\rho$ in the range $\rho_0 - 4\rho_0$ one can
follow the evolution of the polygons. The vanishing of the area of the
polygon at some critical density $\rho_{cr}$ indicates that there is no Skyrme
interaction (having the chosen value of $z$) which can fulfill the chosen
set of constraints at densities beyond $\rho_{cr}$.   

We have performed two types of calculations: a) using only the six
constraints from the Landau parameters of symmetric nuclear matter and
disregarding those of neutron matter; b) using all eleven constraints from
symmetric nuclear matter and neutron matter plus the
sound velocity constraint in neutron matter.
In Fig.~1  
are displayed some typical results. We have
chosen a positive and a negative value of $z$ since there is no {\it a
priori} limitation on $z$. The value $z=2\cdot10^4$ MeV fm$^{3(1+\sigma)}$
correspond to Figs.~1a) and 1b), and $z=-2\cdot 10^4$ to Figs.~1a')
and 1b'). Results involving only Landau parameters of
symmetric matter are displayed
in Figs.~1a) and 1a'), whereas Figs.~1b) and 1b') correspond to results using
all eleven constraints. The contours are drawn every 0.5$\rho_0$ starting
from $\rho_0$.

Let us first examine the results without neutron matter constraints
(Figs.~1a and 1a'). At $\rho=\rho_0$ the contour is a polygon whose upper
and lower horizontal sides are determined by the constraints on $G_0'$ and
$G_0$, respectively, whereas the left and right sides correspond to $F_1'$
and $G_1$. When $\rho$ increases the lower side becomes tilted and the
constraint on $F_0'$ gradually appears as a new side on the right hand side
of the polygon. At $\rho \simeq 2\rho_0$ and above the $F_0'$ constraint is
dominating over the $G_1$ constraint. The surface of the polygon shrinks as
$\rho$ increases. The value of $\rho_{cr}$ is above $4.5\rho_0$. This value
is reached for $x$ and $y$ both negative if $z$ is positive, whereas for $z$
 negative $\rho_{cr}$ corresponds to $x \ge 0$ and $y \le 0$. 

The situation changes somewhat when neutron matter constraints are added
(Figs.~1b and 1b'). At $\rho=\rho_0$ the lower side is given by the
$G_0^{(n)}$ constraint while the right side of the polygon is mainly
determined by the $F_0^{(n)}$ constraint which limits severely the allowed
area. It must be noted that the sound velocity does not bring any limitation
on the ($x,y,z$) parameters in the domain explored here. One can see that
the critical density $\rho_{cr}$ has now a lower value as compared to the
case of Figs.~1a-1a'.

There is no natural limitation to the domain of the $z$
parameter and therefore, we must explore the dependence of the results on
$z$. 
In Fig.~2  
we present calculations where all eleven constraints are
included, exploring the $z$-axis from negative to positive values. The
interpretation of the different sides delimitating the contours is the same
as in the case of 
Fig.~1.  
One can see that $\rho_{cr}$ 
is highest when $z$
is between 0 and $3 \cdot 10^4$. For negative values of $z$ the allowed area at
$\rho=2.5\rho_0$ is already fairly small. Below $z \simeq 3 \cdot 10^4$ the
polygons of larger $\rho$ are contained inside those of smaller $\rho$,
i.e., interactions stable at density $\rho$ are also stable at all
densities between $\rho_0$ and $\rho$. For $z$ larger than $3 \cdot 10^4$ the
allowed areas at larger $\rho$ tend to move outside the area of
$\rho=\rho_0$, which means that that such interactions might seem acceptable
at large densities but they have the defect of having instabilities at
normal density.

So far we have discussed the results obtained with the values of empirical
inputs as adopted in Section 2. It is interesting to see how the conclusions
may depend on this choice. 
In Fig.~3  
we study, for a fixed value of
$\rho=2\rho_0$ and a chosen $z= 10^4$ MeV.fm$^{3(1+\sigma)}$, the
evolution of the allowed polygons when one changes one of the empirical
inputs: the effective mass $m^*_0/m$ (Fig.~3a), the incompressibility $K_0$
(Fig.~3b), the surface energy $\epsilon_S$ (Fig.~3c). It can be seen that
the results are sensitive to the value of $m^*_0/m$. A smaller value of
$m^*_0/m$ makes the allowed domain larger and therefore, the critical
density $\rho_{cr}$ is shifted to higher values. 
On the contrary, the allowed domain is reduced when higher values of $m^*_0/m$ 
are considered, and instabilities appear for densities 
below $2\rho_0$. Hence, to avoid low density
instabilities, one should prefer low $m^*_0/m$.
The sensitivity to $\epsilon_S$ is also non negligible, whereas the dependence
on $K_0$ is moderate.  

In Fig.~4  
we present the calculated values of the critical density
$\rho_{cr}$ (in units of $\rho_0$). Fig.~4a shows $\rho_{cr}$ as a function
of the parameter $z$ for various choices of the empirical input $m^*_0/m$
around the value 0.70~. As pointed out above, one can see clearly that
it is more favorable to choose $z$ positive and less than $3 \cdot 10^4$ 
in order to have $\rho_{cr}$ not too small,
and that larger values of $m^*_0/m$ tend to yield lower $\rho_{cr}$. The
three curves of Fig.~4b represent $\rho_{cr}$ as a function of the relative
variations (in percentage) around the standard choice of $m^*_0/m$, $ K_0$
and $\epsilon_S$. The value of $z$ is fixed at $10^4$ MeV.fm$^{3(1+\sigma)}$.
These curves show that $\rho_{cr}$ depends little on $K_0$ whereas moderate
increases of $m^*_0/m$ (from 0.70 to 0.75), or of $\epsilon_S$ (from 18 MeV
to 20 MeV) can lower $\rho_{cr}$ below 3$\rho_0$.

Now, the question arises whether these bounds for Skyrme parameters
are useful or not
for calculations in pure nucleon matter and finite nuclei. In principle,
we expect a qualitative positive answer for finite nuclei because the used
inputs guarantee that the first terms in a leptodermous expansion of the mass
formula are well described.
To be more quantitative, we have constructed several parametrizations using
the following criteria. We first fix an arbitrary value of $z$, going from
$-10^4$ to $4\cdot 10^4$, and then we take the values of $x$ and $y$
as the coordinates of the point $\Omega(\rho_{cr})$. In this way
instabilities are pushed to the highest values of the density. The sets of
parameters are dubbed Skzn and they are given in Table~\ref{table1}.
Of course,
the criteria of highest value of $\rho_{cr}$ is purely arbitrary but our
intention here is just to explore the resulting parametrizations and
not to give the best parameter set. In Fig.~5 are displayed the
results for pure neutron matter. One can see that choosing negative values
of $z$ results in an increasing neutron effective mass for
increasing densities.
On the other side, values of $z$ greater than $2 \cdot 10^4$ produce too
low binding energies. We have then restricted our calculations in finite
nuclei to parametrizations Skz0, Skz1 and Skz2. In Table~\ref{table2} are
displayed the ground state binding energies and charge radii of
some doubly-magic nuclei
together with the results obtained with SIII and SLy4 interactions.
The results are reasonably satisfactory. We have noticed that convergence
problems occur with Skz0 when looking for the HF solution of $^{208}$Pb.  
To obtain better results
one should relax the condition of highest value for the critical density, and 
of course include some finite nuclei results 
in the fine tuning of the parameters.

\section{Conclusion}

In this work we have explored the parameter domains of Skyrme-type forces
where the stability conditions related to Landau parameters inequalities can
be satisfied up to densities of the order of $4\rho_0$. Stability of both
nuclear matter and pure neutron matter are considered. We have taken
advantage of the possibility to characterize the domains by analytical
expressions. Starting from a general Skyrme force containing 10 parameters,
we have shown that 7 parameters or combinations thereof can be approximately
fixed by physical quantities which can be considered experimentally known to
some extent. Thus, the problem reduces to the study of allowed domains in a
3-dimensional space spanned by 3 parameter combinations.

The results show that, for any Skyrme-type interaction there is a critical
density $\rho_{cr}$ above which one cannot insure all stability conditions.
This critical density does not exceed $3.5-4\rho_0$ for a reasonable choice
of empirical inputs. It is nevertheless an interesting result to know that
it is possible to find Skyrme-type interactions which can give stable
nuclear matter and neutron matter up to such densities. 
The parameter domains are well identified and it would be worthwhile to 
look inside those domains for Skyrme interactions which can also describe 
accurately finite nuclei. Our exploration with parametrizations Skzn has 
shown that reasonable ground state binding energies and charge radii 
can be obtained. A more systematic
study over a wide number of nuclei would be necessary to get the Skyrme-type
interaction which is stable over the largest range of densities. This new
interaction would be very useful for neutron star calculations, like for
instance the neutrino mean free path or the URCA process.


The present analysis is restricted to Skyrme interactions, because the
simplicity of the resulting expressions allows for an algebraic
study. The presence of instabilities in nuclear matter and neutron matter 
beyond some critical density could be attributed to the  zero-range
form of the interaction, and its $k^2$-dependence of the effective masses. 
Finite range effective interactions~\cite{gog75} are an alternative choice 
for describing the equation of state of nuclear and neutron matter.
We know that the actual parametrizations (D1, D1S, D1P)
do not predict instabilities~\cite{mar01} for densities below $\simeq 5\rho_0$.
We have not considered a systematic study of Gogny effective interactions
because they contain more parameters, and they do not allow 
for a simple algebraic analysis. 

\subsection*{Acknowledgments}
We have greatly benefited of fruitful discussions with our late colleague
Dominique Vautherin on this subject and we acknowledge his invaluable help.
This work has been partly supported by DGI (Spain), grant BFM2001-0262. J.N.
and N.V.G. thank GANIL for its hospitality during completion of the work.

\appendix

\section{Nuclear matter relations}
Employing a Skyrme interaction, the 
energy density of semi-infinite symmetric nuclear matter
is written as
\be
{\cal H} = \frac{\hbar^2}{2m^*} \tau + \frac{3}{8} t_0 \rho^2 +
\frac{1}{16} t_3 \rho^{\sigma+2} + \frac{1}{8} T_S | \nabla \rho |^2
- \frac{m^*}{\hbar^2} V_{\rm so} \rho | \nabla \rho |^2~,
\label{A1}
\ee
where 
$V_{\rm so}= 9 W_0^2/16$. 

Denoting by $\rho_0$, $\epsilon_0$, $K_0$ and $m^*_0$ respectively
the density, energy per particle, compressibility, and effective mass
at saturation of the bulk symmetric nuclear matter, it turns out
that the Skyrme parameters $t_0$, $t_3$, $\sigma$, and the
combination $T_0$ can be written as:
\bea
&&T_0 = \left( \frac{\hbar^2}{m^*_0} -
\frac{\hbar^2}{m} \right) \frac{1}{\rho_0}~, 
\label{A2}
\eea
\bea
&& \sigma = \frac{\ds \frac{1}{9} K_0 + \epsilon_0 + \left(
\frac{\hbar^2}{10m} - \frac{2 \hbar^2}{15m^*_0} \right) k_F^2(0)}
{\ds - \epsilon_0 + \left(\frac{3 \hbar^2}{10m} -
\frac{\hbar^2}{5m^*_0} \right) k_F^2(0)}~,  
\label{A3}
\eea
\bea
&&t_3 = \frac{16}{\rho_0^{1+\sigma}} \frac{1}{\sigma}
\left[- \epsilon_0 + \left(\frac{3 \hbar^2}{10m} -
\frac{\hbar^2}{5m^*_0} \right) k_F^2(0) \right]~,  
\label{A4}
\eea
\bea
&&t_0 = \frac{8}{3 \rho_0} \left[ \epsilon_0 - \frac{3\hbar^2}{10
m^*_0} k_F^2(0) - \frac{1}{16} t_3 \rho_0^{1+\sigma} \right]~, 
\label{A5}
\eea
where $\rho_0 = 2/(3\pi)^2\,k_F^3(0)$. 
Once the parameters $t_0, t_3, \sigma$ and the combination $T_0$
have been fixed, the surface energy only depends on the combination 
$T_S$ and the spin-orbit strength $W_0$. The surface energy 
can be written as an integral over the density~\cite{trekri}
\bea
 \epsilon_S & = & 8 \pi r_0^2  \int_0^{\rho_0} d\rho 
\left[ \frac{\hbar^2}{36m} - \frac{5}{36} T_0 \rho + \frac{1}{8}
T_S \rho - \frac{m^*}{\hbar^2} V_{\rm so} \rho^2 \right]^{1/2} 
\\
\label{A6}
 &  & \hspace{2cm}
\left[\frac{3 \hbar^2}{10m^*} k_F^2 + \frac{3}{8} t_0 \rho+
 \frac{1}{16} t_3 \rho^{\sigma+1} \right]^{1/2}~, \nonumber
\eea
where $r_0 = \left[ 3 / 
(4 \pi \rho_0) \right]^{1/3}$ is the unit radius, and a Thomas-Fermi 
approximation up to $\hbar^2$-order has been used to replace
the kinetic energy density.

Then, the following relation involving the symmetry energy $\epsilon_I$ can
be used to relate $x_0$ to the already known parameter combinations and to
the free parameters $(x,y,z)$:
\be
t_0 x_0  =  \left( - \frac{3}{2} x + \frac{9}{10} y + \frac{2}{5} (T_0 - 2
T_S)\right) \alpha_1 \rho_0^{2/3}  
 - \frac{1}{6} z \rho_0^{\sigma} - 2 C_1(\rho_0) - 
 \frac{4 \epsilon_I}{\rho_0}~. \label{A12}
\ee

To simplify the notation used in Section 2 we have introduced the 
following functions of the density:
\bea
C_0(\rho) &=& \frac{\hbar^2}{m} + T_0 \rho~, 
\label{A7}
\eea
\bea
C_1(\rho) &=& - \left( \frac{\hbar^2}{m} + T_0 \rho \right)
\frac{\alpha_1}{\rho^{1/3}} + \frac{1}{4} t_0 +
\frac{1}{24} t_3 \rho^{\sigma}~, 
\label{A8}
\eea
\bea
C_2(\rho) &=& - \frac{1}{2} \left( \frac{\hbar^2}{m}+ 4 T_0 \rho \right)
\frac{\alpha_2}{\rho^{1/3}} - \frac{1}{2} t_0 -
\frac{1}{24} (\sigma+1)(\sigma+2) t_3 \rho^{\sigma} 
\label{A9}
\eea
\bea
C_3(\rho) &=& - \frac{1}{2} \left( \frac{\hbar^2}{m} + T_0 \rho \right)
\frac{\alpha_2}{\rho^{1/3}} + \frac{1}{2} t_0 +
\frac{1}{12} t_3 \rho^{\sigma}~, 
\label{A10}
\eea
which contain only known parameters. We have also introduced the constants
\be
\alpha_1 = \left(\frac{\pi^4}{12} \right)^{1/3} \quad , \quad
\alpha_2 = \left(\frac{\pi^4}{3} \right)^{1/3}
\label{A11}
\ee 

\section{Landau parameters}
The Landau parameters in symmetric nuclear matter are:
\bea
F_0 &=& \left( \frac{3}{4} t_0 + \frac{1}{16} (\sigma+1)(\sigma+2) t_3
\rho^{\sigma} \right) \frac{2m^* k_F}{\hbar^2 \pi^2} -F_1~, 
\label{B1}
\eea
\bea
G_0 &=& \left( \frac{1}{4} t_0 (2x_0-1) + \frac{1}{24} t_3
\rho^{\sigma} (2x_3 -1) \right) \frac{2m^* k_F}{\hbar^2 \pi^2} -G_1~,
\label{B2}
\eea
\bea
F'_0 &=& \left( - \frac{1}{4} t_0 (2x_0+1) - \frac{1}{24} t_3
\rho^{\sigma} (2x_3 +1) \right)  \frac{2m^* k_F}{\hbar^2 \pi^2} -F'_1~,
\label{B3}
\eea
\bea
G'_0 &=& \left( - \frac{1}{4} t_0 - \frac{1}{24} t_3 \rho^{\sigma} 
\right)  \frac{2m^* k_F}{\hbar^2 \pi^2} -G'_1~,
\label{B4}
\eea
\bea
F_1 &=& - 3 T_0 \frac{m^*}{\hbar^2} \rho~,
\label{B5}
\eea
\bea
G_1 &=& - 3 T_1 \frac{m^*}{\hbar^2} \rho~,
\label{B6}
\eea
\bea
F'_1 &=&  3 T_2 \frac{m^*}{\hbar^2} \rho~,
\label{B7}
\eea
\bea
G'_1 &=&  3 T_3 \frac{m^*}{\hbar^2} \rho~,
\label{B8}
\eea
where the $T_i$'s are the following parameter combinations: 
\bea
T_0 &=& \frac{1}{8} \left[ 3 t_1 + 5 t_2 + 4 y \right]~,
\label{B9}
\eea
\bea
T_1 &=& \frac{1}{8} \left[ 2 x + 2 y - t_1 + t_2  \right]~,
\label{B10}
\eea
\bea
T_2 &=& \frac{1}{8} \left[2 x - 2 y + t_1 - t_2  \right]~,
\label{B11}
\eea
\bea
T_3 &=& \frac{1}{8} \left[ t_1 - t_2 \right]~,
\label{B12}
\eea
and the effective mass is:
\be
\frac{\hbar^2}{m^*} = \frac{\hbar^2}{m} + T_0 \rho~.
\label{B13}
\ee

The Landau parameters in neutron matter are:
\bea
F^{(n)}_0 &=& \left( \frac{1}{2} t_0 (1-x_0) 
+ \frac{1}{24} (\sigma+1)(\sigma+2) t_3
\rho^{\sigma} (1-x_3) \right)  \frac{m^*_n k_F}{\hbar^2 \pi^2} -
F^{(n)}_1~,\\
G^{(n)}_0 &=& \left( \frac{1}{2} t_0 (x_0-1) + \frac{1}{12} t_3
\rho^{\sigma} (x_3 -1) \right) \frac{m^*_n k_F}{\hbar^2 \pi^2} -
G^{(n)}_1~, \\
F^{(n)}_1 &=& - 3 (T_0 -T_2) \frac{m^*_n}{\hbar^2} \rho~, \\
G^{(n)}_1 &=& - 3 (T_1 -T_3) \frac{m^*_n}{\hbar^2} \rho~, 
\label{B17}
\eea

and the effective mass is:
\be
\frac{\hbar^2}{m^*_n} = \frac{\hbar^2}{m} + (T_0 - T_2) \rho~.
\label{B18}
\ee

\newpage

\begin{table}[htb]
\centering
\begin{tabular}{|c|c|c|c||c|c|c|c|c|}
\hline
 & $z$ & $x$ & $y$ & $t_2$ & $x_0$ & $x_1$ & $x_2$ & $x_3$ \\ 
 & MeV fm$^{3(1+\sigma)}$ & Mev fm$^5$ & MeV fm$^5$ & MeV fm$^5$ & & & & \\
\hline
Skz-1 & $-10^4$ & 570.42 & 266.2 & -299.14 &-0.2665 & 1.2968 & -0.8899 &-0.7282 \\
Skz0 & $0$ & 458.0 & 215.0 & -258.18 & 0.1986 & 1.0413 & -0.8328 & 0.0   \\
Skz1 & $10^4$ & 217.0 &  221.0 &  -262.98 & 0.6052 & 0.4933 & -0.8404 & 0.7282\\
Skz2 & $2 \cdot 10^4$ & 24.0 & 227.0 & -267.78 & 1.0290 & 0.0546 & -0.8478 & 1.4564 \\
Skz3 & $3 \cdot 10^4$ & -265.5 & 233.1 & -272.66 & 1.4174 &-0.6082 & -0.8549 & 2.1846\\
Skz4 & $4 \cdot 10^4$ & -512.0 & 239.15 & -277.50 & 1.8226 &-1.1640 & -0.8618 & 2.9127 \\
\hline
\end{tabular}
\caption{Skzn parametrizations. All of them have the  
following common values: $\sigma$=0.1694, 
$t_0 = -2471.10$~MeV~fm$^3$, $t_1 = 439.85$~MeV~fm$^5$, 
$t_3 = 13732.8$~MeV~fm$^ {3(1+\sigma)}$, as they come from the 
initial inputs.}
\label{table1}
\end{table}

\begin{table}[htb]
\centering
\begin{tabular}{|c|c|c|c|c||c||c|c|}
\hline
Nucleus & Observable & Skz0 & Skz1 & Skz2 & Exp. & SIII & SLy4\\ 
\hline
$^{16}$O   & B/A(MeV) & 8.39 & 8.38 & 8.38 & 7.98 & 7.95 & 7.97 \\
         & $r_c$ (fm) & 2.76 & 2.76 & 2.76 & 2.73 & 2.76 & 2.80 \\
$^{40}$Ca  & B/A(MeV) & 8.87 & 8.85 & 8.85 & 8.55 & 8.48 & 8.55 \\
         & $r_c$ (fm) & 3.47 & 3.47 & 3.47 & 3.49 & 3.50 & 3.51 \\
$^{208}$Pb & B/A(MeV) & N.C. & 7.97 & 7.91 & 7.87 & 7.79 & 7.80 \\
         & $r_c$ (fm) & N.C. & 5.48 & 5.49 & 5.50 & 5.58 & 5.52 \\
\hline
\end{tabular}
\caption{HF binding energies and charge radii for several 
  Skzn parametrizations.
N.C. means that the HF calculation has not converged.}
\label{table2}
\end{table}

\newpage

\begin{figure}[htb]
\label{figure1}
\centering
\includegraphics[scale=0.6]{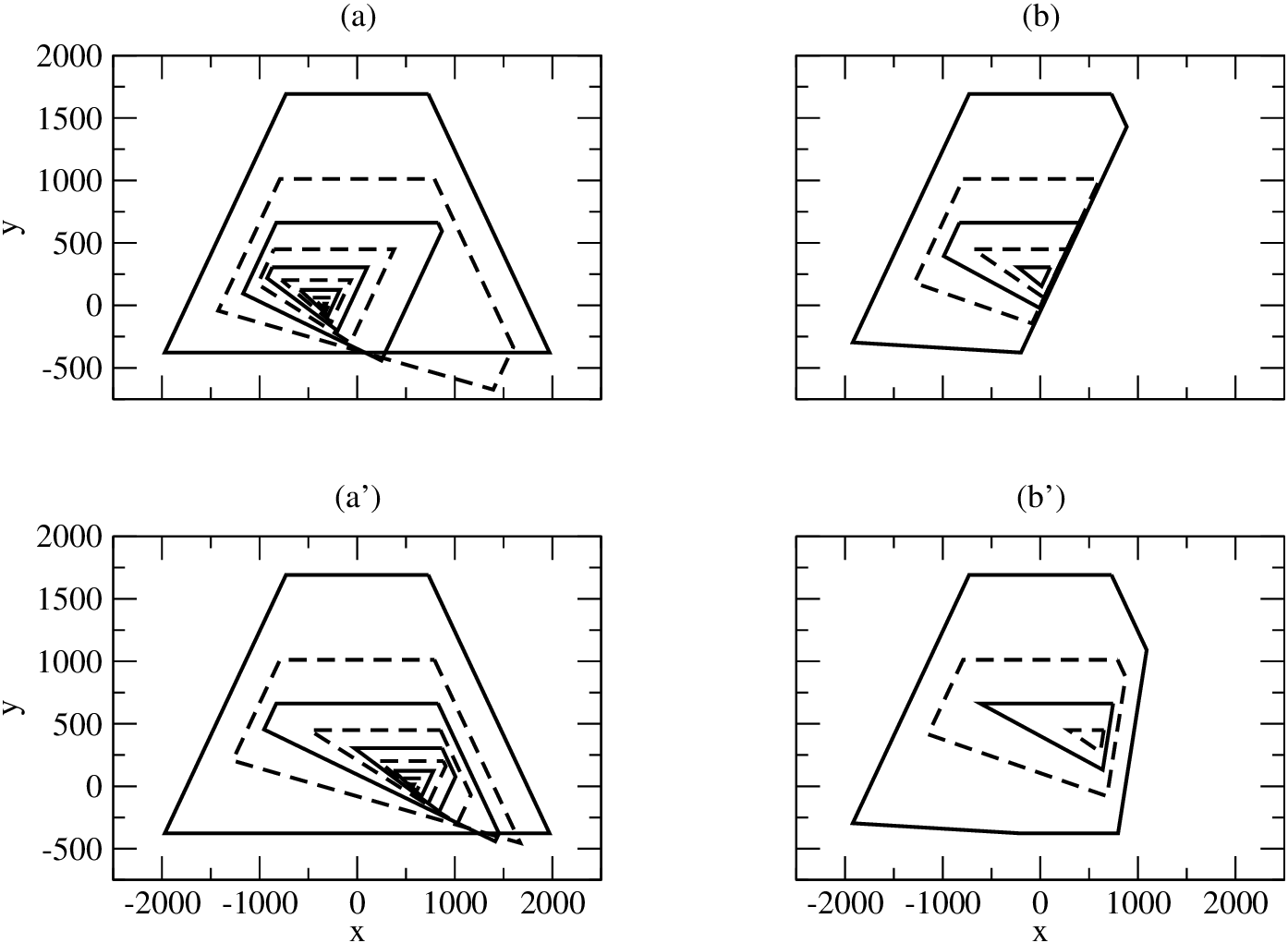}
\vspace{1pt}
\caption{Comparison of the sections of the volume $\Omega(\rho)$ by a
horizontal plane $z$ = constant, in units of MeV fm$^{3(1+\sigma)}$. 
The horizontal and vertical axis are for the parameters $x$ and $y$, 
respectively, in units of MeV fm$^5$. Cases a) and a') correspond to the 
constraints  from the Landau parameters of symmetric nuclear matter.
Cases b) and b') include also the constraints from neutron matter.
Two values of $z$ have been used, namely $z=2\cdot 10^4$ for cases a) 
and b), and $z=-2\cdot 10^4$ for cases a') and b'). 
The different closed contours correspond to different values of $\rho$. 
The largest area is for $\rho=\rho_0$, the next one corresponds to an 
increase by a step of 0.5$\rho_0$, and so on. For the sake of clarity, 
they are alternatively represented by solid and dashed lines.}
\end{figure}

\newpage*

\begin{figure}[htb]
\label{figure2}
\centering
\includegraphics[scale=0.6]{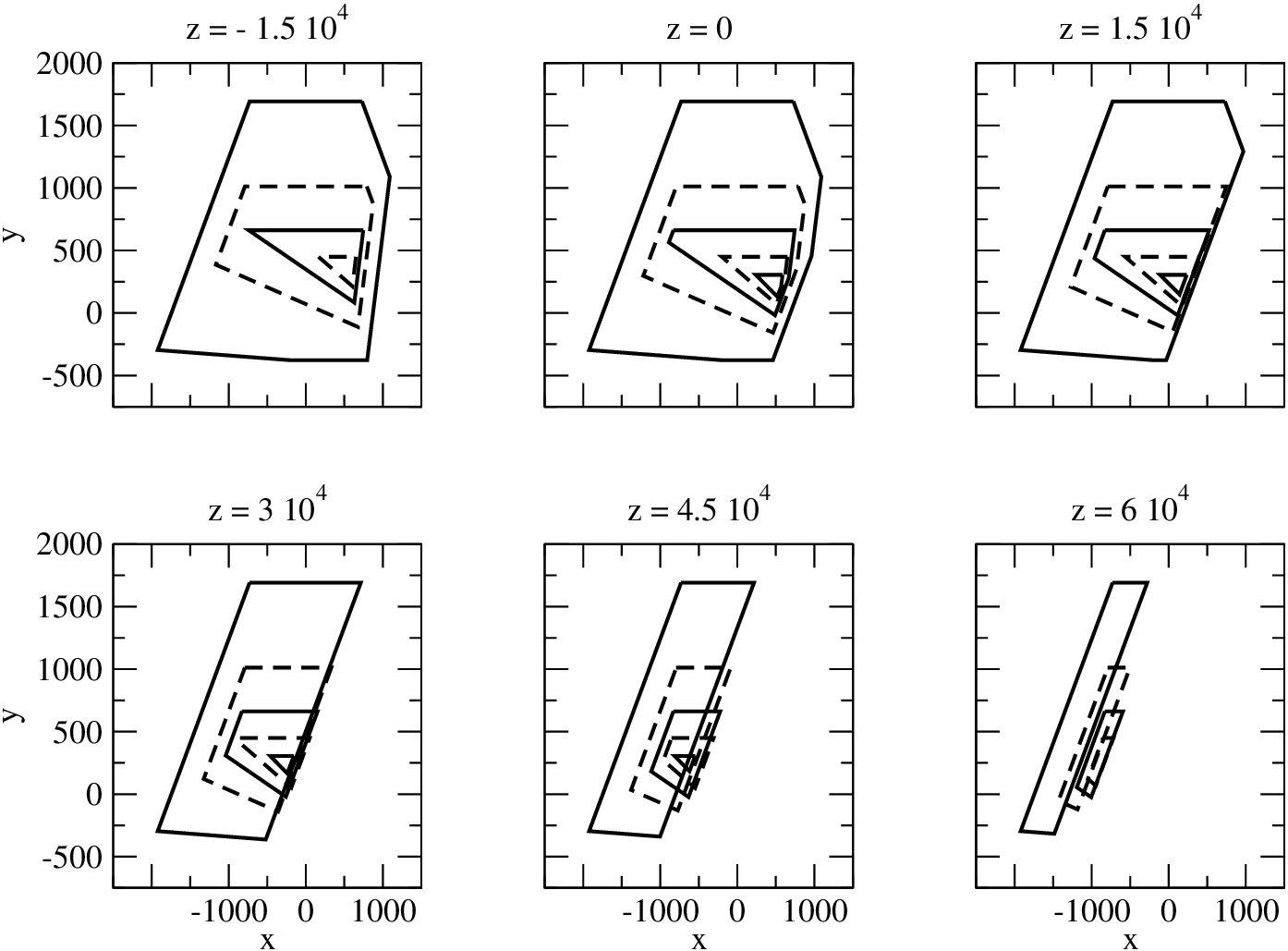}
\vspace{1pt}
\caption{Comparison of the sections of the volume $\Omega(\rho)$ by a
horizontal plane, for choices of $z$ different from Fig.~1. The various closed
contours correspond to different values of $\rho$. The largest area is
for $\rho=\rho_0$, the next one corresponds to an increase by a step of
0.5$\rho_0$, and so on. The calculations are done using the full
constraints. The values of $z$ are in units of MeV.fm$^{3(1+\sigma)}$.}
\end{figure}

\newpage*

\begin{figure}[htb]
\label{figure3}
\centering
\includegraphics[scale=0.6]{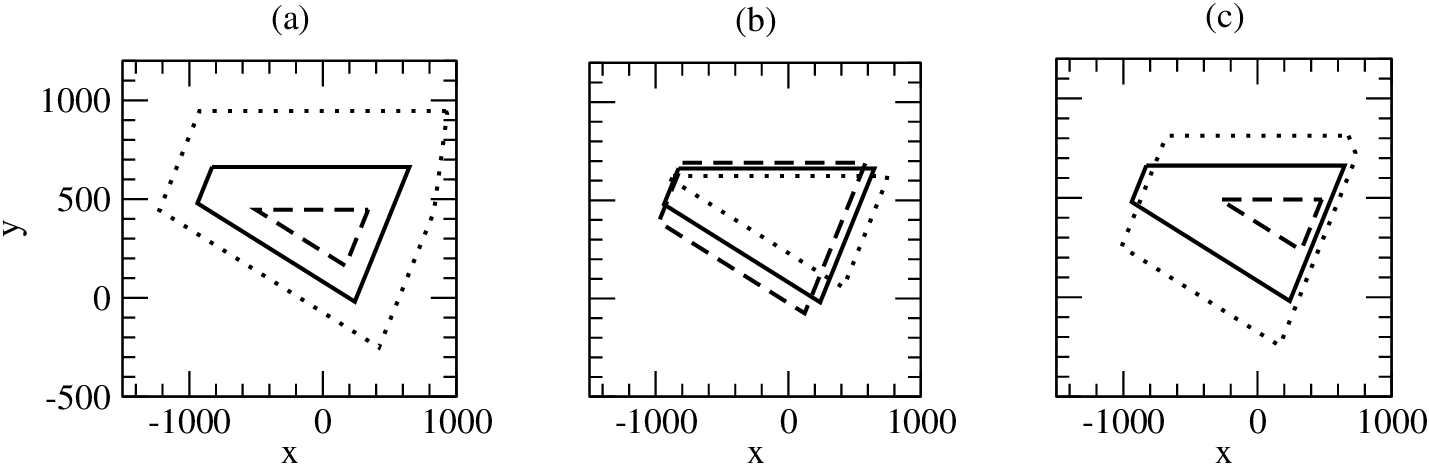}
\vspace{1pt}
\caption{Sensitivity to some empirical inputs. Case a:
results for $m^*_0/m=$ 0.6, 0.7 and 0.8 (respectively dotted, solid, dashed), 
keeping the remaining empirical inputs to their fixed
values. Case b: same as a), with
$K_0=$ 210, 230 and 250 MeV
(respectively dotted, solid, dashed). Case c: same as a),
with the surface
energy $\epsilon_S$= 16, 18 and 20 MeV (respectively dotted, solid, dashed). 
All cases have been calculated for 
$z=10^4$ MeV fm$^{3(1+\sigma)}$ and $\rho=2\rho_0$.}
\vspace{1pt}
\end{figure}

\newpage*

\begin{figure}[htb]
\label{figure4}
\centering
\includegraphics[scale=0.6]{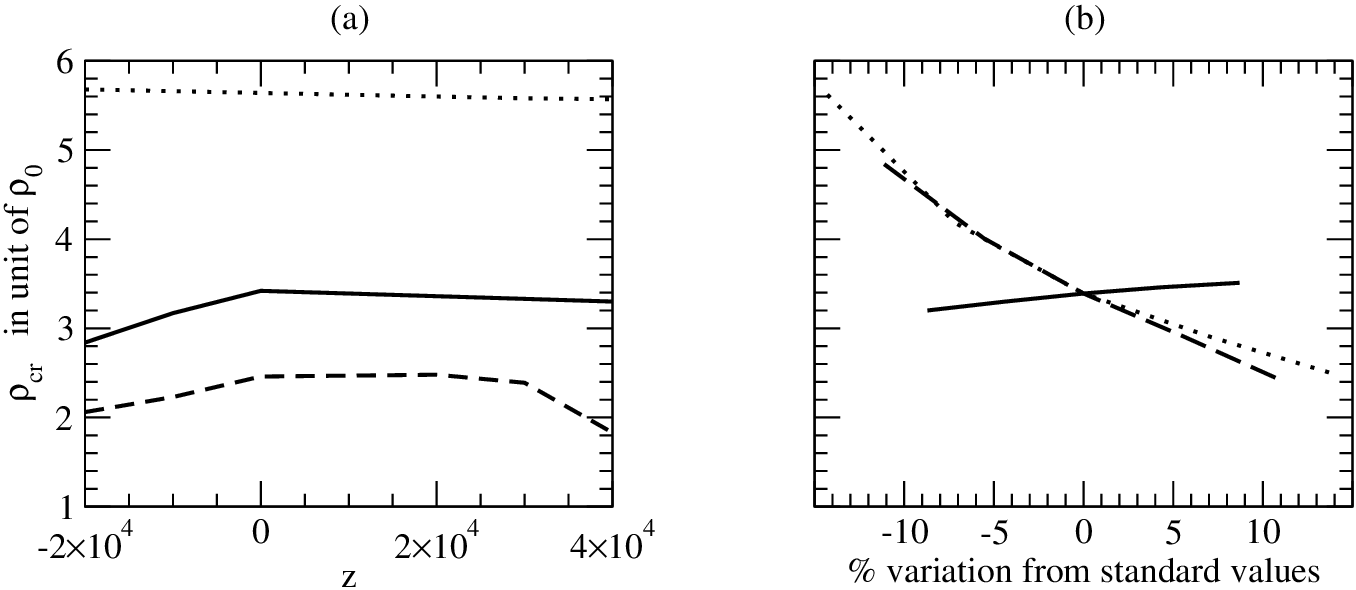}
\vspace{1pt}
\caption{Case a: $\rho_{cr}$ as a function of $z$ (in units of
  MeV.fm$^3(1+\sigma)$) for different values of $m_0^*/m$=0.6,0.7,0.8
  (respectively dotted, solid, dashed). 
  Case b:
  $\rho_{cr}$ as a function of relative variations of some empirical inputs;
  dotted line: variation of $m_0^*/m$, solid line: variation of $K_0$,
  dashed line: variation of $\epsilon_S$.   }
\end{figure}

\newpage*

\begin{figure}[htb]
\label{figure5}
\centering
\includegraphics[scale=0.6]{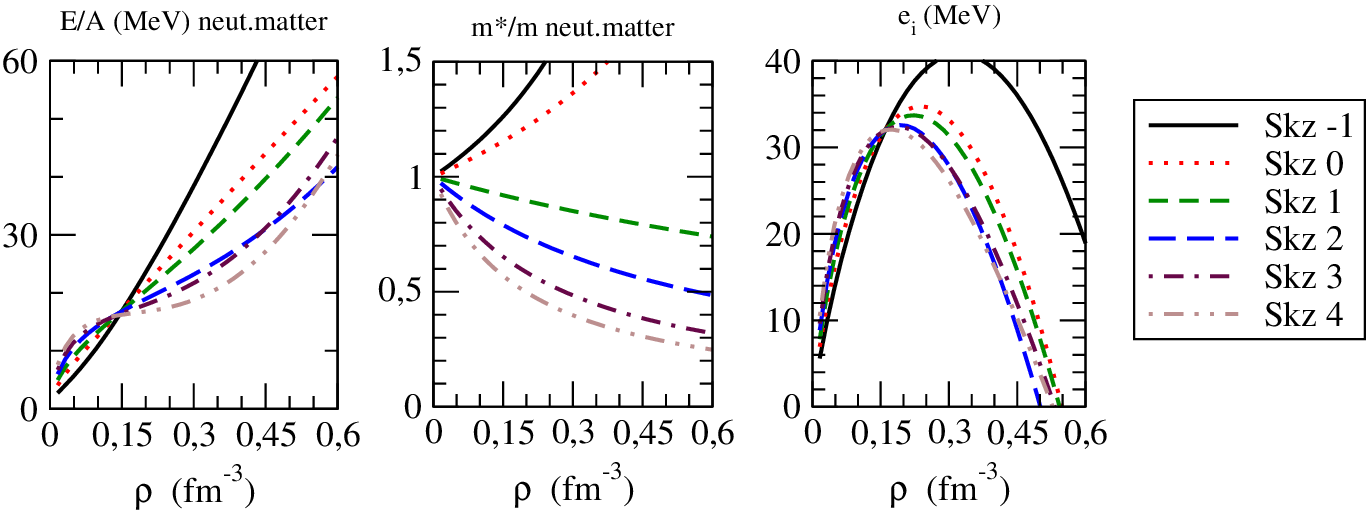}
\vspace{1pt}
\caption{Predictions of several Skzn parametrizations for the
binding energy and the effective mass of pure neutron matter and for
asymmetry energy as function of the density.
}
\end{figure}

\end{document}